# DYNAMIC COMPONENT COMPOSITION

Efim Grinkrug[1]

[1]Department of Software Engineering, National Research University Higher School of Economics (HSE), Moscow, Russia

## ABSTRACT

*This paper presents an approach to dynamic component composition that facilitates creating new composed components using existing ones at runtime and without any code generation. The dynamic abilities are supported by extended type notion and implementation based on additional superstructure provided with its Java API and corresponding JavaBeans components. The new component composition is performed by building the composed prototype object that can be dynamically transformed into the new instantiable type (component). That approach demonstrates interrelations between prototype-based and class-based component-oriented programming. The component model proposed can be used when implementing user-defined types in declarative languages for event-driven applications programming.*

## 1.    INTRODUCTION

Component-oriented programming (COP) is a programming paradigm that enables programs to be constructed from prebuilt software components, which are reusable, self-contained blocks of computer code [1]. It provides many advantages from various points of view in software development process and constitutes the main idea of component-based software engineering (CBSE).

The idea of component-oriented programming is to create software products from composing parts – the idea that is at the base of the vast majority of technologies in other engineering areas. Almost everything is made from components in industry, even "monolithic" products may be created using components (e.g., to make casting molds for them).

Composing parts of software products named components are created and used in accordance with a component model that defines what a component is, and what and how can be composed with that component [2]. Plenty of various component models were proposed for different platforms, application areas and purposes.

In this paper we use Java VM as an implementation platform and consider the most popular component model for that platform – JavaBeans component model – as an initial point of our discussion. Java platform has become the most widely used object-oriented environment for software development starting from the time it was introduced [3]. JavaBeans component model [4] initially was claimed as "the only component model for the Java machine" and it is widely used up to now, while currently there are many other, popular enough component models for Java-platform – both universal and domain specific [5].

According to the JavaBeans component model, JavaBeans - component is a class that can be instantiated in any context and has support for persistence. The component produces the component instances by its public default constructor (having no arguments). The component instances can be composed with each other by referencing one instance from another and resulting in a graph of component instances that can be built dynamically and interactively having a corresponding builder tool support (i.e., without having a compiled program to build that graph, while it is possible in traditional Java-programming).

However, using standard JavaBeans - components for composition in that way, we cannot produce a new composed JavaBeans – component (as the composition result) dynamically, without compilation (or JVM code generation). In order to do that, we need to generalize the notion of component and enhance the JavaBeans component model accordingly; that is the goal of this work.

Initially, this work was rooted from the practical project [6] where JavaBeans – components were used to implement a subset of Virtual Reality Modeling Language (VRML) [7] with its 3D-graphics support entirely in Java. The experience earned helps to highlight some drawbacks of the JavaBeans component model along with shortcomings of the VRML specification, especially concerning user-defined components definition and implementation. We expect that results of this work may be general enough and useful when developing applications with event-driven behaviours in component-based manner.

We begin with the problem statement, discussing how components are used for compositions and what the drawbacks of the JavaBeans component model are in Section 2, explaining our approach to the component model enhancements. Then we describe our component model and its implementation principles in Section 3. After that in Section 4 we discuss how that component model can be used in various applications. We find its place in the wide variety of component-based software technologies in Section 5 and conclude with future work directions.

## 2. COMPONENT COMPOSITIONS

Component-oriented programming and software architecture in general deal with component compositions. These notions should be properly defined. Since the time when component-oriented programming was recognized as the base of software engineering [8], many definitions of a component were introduced. The most popular are following:

- "A component is a software element (modular unit) satisfying the following conditions:
1. It can be used by other software elements, its 'clients'.
2. It possesses an official usage description which is sufficient for a client author to use it.
3. It is not tied to any fixed set of clients."[9];
That definition is the most general, emphasizing the role of a component for compositions.
- "A software component is a unit of composition with contractually specified interfaces and explicit context dependencies only. A software component can be deployed independently and is subject to composition by third parties" [10];
That definition explicitly states that component is subject to composition, but, in our concrete point of view, hides some important details on what is actually deployed and composed.

- "A [component is a] software element that conforms to a component model and can be independently deployed and composed without modification according to a composition standard."[11].

That definition seems to be more specific: it refers component model and composition standard for future refinements – what specifically and how should be composed.

A software component model mentioned in the last definition, should define the syntax of components (how they are constructed and represented), the semantics of components (what components are meant to be), and the composition of components (how they are composed or assembled) [2]. A component model and composition standard may be dependent on platform, composition and execution environment. We are specifically targeting Java platform and consider JavaBeans component model as subject for investigation.

## 2.1. Components in Object-Oriented Environment

While object-oriented environment is not necessary for component-oriented programming, it can benefit from using object-oriented environment as its base component implementation and runtime platform. In general, object-oriented environment emphasizes modularity of the construction of the system, while component-oriented development emphasizes design, production, deployment and use of the system [12], i.e., aspects of the system development that are more dynamic and less dependent on static tools (like compiler, etc.). Object-oriented design emphasizes the development time relationships between entities in a system before a system is built using static tools, while component-oriented design emphasizes more the dynamic (deployment and runtime) phases.

Generally speaking, when coding in Java language we can only write data types definitions and implementations that are then compiled into class-files with byte-codes and appear to be the runtime types (instances of java.lang.Class class) in JVM upon class loading. We cannot define an individual object (instance) without having its class. Java-programming corresponds to class-based programming paradigm; we can only define types of objects (classes) to be instantiated [13]. In that sense we can consider Java platform as the *implementation environment* for components that are developed using traditional static development tools (compiler or, possibly, other byte-code generators, like BCEL [14], used for generative programming, i.e., automation of code development [12]). Types (classes) and components (types implemented by classes of special kind) developed with target platform code generation (using these static tools) we name as *hardcoded* types and components, correspondingly.

According to the JavaBeans component model, JavaBeans-components are implemented by Java classes that satisfy simple rules (JavaBeans design patterns). Essentially, JavaBeans- component is a Java class instantiable in any context (public class having public constructor with no arguments) and having support for persistence (to save and restore states of its instances). These classes are distributable for reuse in binaries (Java byte-codes) along with other classes and resources used to implement them. A JavaBeans-component, therefore, is a Java class that implements, possible using other Java classes, a type of objects, or instances, it creates; component is not that instance. These notions are often erroneously mixed in literature hiding class-based object-oriented nature of the component model. Components are distributed and deployed in form of classes (possibly, with other resources).

## 2.2. JavaBeans – component compositions

At the design stage, components composition can be performed using Java programming technology chain, resulting in components byte-codes produced by compiler that actually uses the composing components, if needed, just like other classes from class libraries. The goal of JavaBeans component model development was to benefit from the fact that Java-platform is more than just object-oriented programming language, but provides dynamic abilities for component composition.

The popularity of the JavaBeans component model is based on its relative simplicity, wide range of abilities it provides and corresponding tools to demonstrate these abilities, especially dynamic ones. The first sentence of the specification is: "The goal of the JavaBeans APIs is to define a software component model for Java, so that third party ISVs can create and ship Java components that can be composed together into applications by end users" [4]. The expected scenarios for JavaBeans-components composition differ in their levels of supported dynamics. While JavaBeans-components can be used as class libraries in traditional software development process, we are mainly interested in their dynamic composition abilities and correspondent tools support.

From an external view, JavaBeans-components offer four kinds of ports for their instances communication: methods, properties, event sources and event listeners.

The notion of method of a component is directly bound to the notion of method in the component implementation language: callable method must be present in the component implementation class that must be compiled into byte-codes and loaded.

The notion of property, from a client point of view, can be used for getting its value, setting a new value, or for binding with it to be notified with property change events whenever it changes its value. That notion is supported by JavaBeans property design pattern and corresponding dynamic introspection mechanism based on reflection.

Event sources generate events of certain types, while event listeners receive the events. Event sources provide operations to connect and disconnect listeners, supporting event-driven behaviours in applications composed of components. Binding events to listeners requires using some generative programming (to create event hook-ups) or dynamic proxy mechanism (a kind of internal generative programming embedded into JVM for that purposes). Binding property change events does not require code generation having dynamic reflection-based support.

In component-based software engineering, composition is a central issue, since components are supposed to be used as building blocks from some repository and assembled or plugged together into larger blocks or systems.

JavaBeans-component model can be considered from a general, idealized component model point of view [2, 15, 16], that is expected to have three stage lifecycle: 1) design stage, when components are designed and developed at source code level of their implementation language (i.e., in Java, in our case) and possibly compiled into binaries; that is the stage of hardcoded components creation; 2) deployment stage, when binary component representations are supplied into composition environment, and 3) runtime stage, when the components are instantiated and the instances are working in a running program. Actually, what we are going to do is remove borders between the stages.

At the design stage components can be used by compiler as classes from class library; it means that compiler can perform *static component composition*, while we are looking for *dynamic component composition* without code generation (without generative programming support).

At the deployment stage, some composition environment that supports JavaBeans composition visually is expected. The JavaBeans API Specification was supplied with Bean Development Kit (BDK [17]) that contained the composition environment prototype, the BeanBox, to illustrate interactive and visual composition support for JavaBeans components. That approach has been integrated in various IDEs, but the BeanBox from BDK is still used having more dynamic abilities than the IDEs provide. At that stage, JavaBeans components are instantiated in the composition environment and their instances are combined together to provide the composite functionality required. In the BeanBox, components instances can be combined interactively and dynamically using all kind of ports to connect them together. When component instances are composed in the BeanBox tool, the whole composition can be saved (serialized using binary or some other format), and restored (de-serialized) later.

In contrast with the BeanBox, IDEs use more static approach: they do not support direct component instances interactions, but help generating the source code for it, that gets compiled after the generation. In that sense, we can consider that IDEs as the design stage tools that use composition ability to automate some code generation.

At the runtime stage, components that were created at the design stage are instantiated and executed. But some of them, acting as the containers can use serialized composite objects stored at the design stage by de-serializing them inside their instances. In any case, we see that our JavaBeans composition abilities could not create a new component without compiling it. We cannot produce a new component in BeanBox-like interactive tool: components must be classes, and classes may be created by their byte-codes generation only.

At the same time, the visual composition of pre-existing components is at the origin of the JavaBeans component model and it is stated in the initial definition: "A JavaBean is a reusable software component that can be manipulated visually in a builder tool" [4]. And the abilities of JavaBeans components to be manipulated visually (i.e., dynamically) make JavaBeans component model attractive to use when developing various modelling and visualizing applications with interactive abilities support.

## 2.3. Class-based and Prototype-based Component-Oriented Programming

The advantage of the JavaBeans component model is in its simplicity and usability: it is not based on a set of specific interfaces to be implemented by the components (statically, by corresponding code generation), as e.g. OSGi does [5]. Enhancements of the JavaBean component model that had introduced BeanContext related features in that style are much less popular.

Component-based technologies are used in many other engineering areas. In most of them instruments used to create their basic components differ from instruments used to build composed components from them. Usually, composing technologies are simpler (remind, e.g., so called "screwdriver production" for computers).

In software, XML syntax is often used to declaratively define and create a composite object from components instances; that XML-based instance composition is much more simple technology than the compiler used to create the components (classes) themselves. For instance, we used that technology in [6] to compose VRML-scene [7] from JavaBeans-components instances in the same way that is used, e.g. in XAML [18], but we used the VRML-parser of our own. These instruments to build a composed object from components instances are significantly simpler than Java compiler required to produce JavaBeans-components. Moreover, building a composite object can be done interactively, as the Bean Box demonstrates for JavaBeans-components.

When we build a composite object from JavaBeans-components instances using some declarative language, or when we build the composite object from JavaBeans-components instances in a composer tool (similar to the Bean Box), we discover that the component technology we use is not "self-closed".

Our source components are types of objects implemented as classes (compiled and supplied in form of instantiable types libraries). As a result of composition made with a parser or interactively in a tool we get some composite object composed from supplied components instances or an instance of some predefined container type filled with our composite object.

A composite object we have created may be a workable object (e.g., it can be some GUI implementation or a 3D-scene made from VRML subset implemented, like in [6]). We can save and restore that composite object, we can clone it (either directly or by serialization/deserialization), but we always deal with it as an instance of some (predefined) container type, but not as a new (component) type, that can be used the same way as the components that were used to compose it. Having the components set in hand initially, we cannot produce some new composed component that can be instantiated like a type, instead of cloning it like composed object.

It is graphically visible while manipulating in BeanBox. First, a library with compiled JavaBeans-components classes is loaded into BeanBox repository (named as ToolBox). In the BeanBox container instance we instantiate components (types) that are dragged & dropped into it from the ToolBox. The instances just created are depicted inside BeanBox container instance, having their property values shown at the Properties panel, where some values can be edited. Further on, we can link the instances together by their references, assigning one instance as some property value for another, or bind them by events. All that BeanBox provided abilities correspond to the JavaBeans Specification [4], that the tool was aimed to illustrate. We just want to highlight the lack of ability to enrich the ToolBox, filled with components we used initially, with the result of our manipulations with them during the composition.

The technology we use is not "self-closed" since by simple manipulations with components we cannot define composed component that can be manipulated the same way.

It is important to note that we would like to have that ability by simple means, i.e. by means other than the means that were used to create initial components. (Compare: electronic LSI components are created by more sophisticated technology than the one used for putting them on a PCB together.)

Initially, basic JavaBeans-components are represented as compiled classes that were conformed to the JavaBeans-component definition and the JavaBeans design patterns while they were coded. Creating JavaBeans-components is usually done statically as the result of standard software developing process, with packaging their byte-codes into a Jar-archive. We say "usually" because some dynamic code generation is possible (either by creating source code with its compilation on the fly, or by mean of immediate code generation using specific libraries for that purpose that are available). Both variants can be used at runtime (dynamically), but we do not consider them as "simple".

Basic JavaBeans-components are created statically, by "hand-made" programming. When we use "hand-made" (or even automated to some extent) programming while creating new composed JavaBeans-component, using existing JavaBeans-components just as class libraries, then we apply the same technology that was used to produce basic components initially.

While that approach can be (and is) widely used (and, being sophisticated, can provide more effective result), it seems that a way to create composed component by means other than that of basic ones, can give some advantages – both technical and ideological.

From the technical point of view, the ability to define composed component interactively (and without compiling, dynamically) is attractive by simplicity of its usage – just as putting LSI circuits on the PCB (while we can have VLSI circuit for all of them later).

From the ideological point of view, we can talk about creating some "higher virtualization level" with its new type system, incorporating the lower level abilities to create and support base components functioning (i.e., on top, or above the Java-platform). Base JavaBeans-components are represented as instantiable types (implemented as Java classes). In case we want to get new composed component "by means of simple manipulations" (i.e., without code generation), we need to generalize a component (type) notion so that we can, when creating new composed types, use basic and composed components (types) in the same manner (equally). Along this way, compiled (or hardcoded) components and composed components are just two kinds of type implementations at our "new virtualization level". An idea of that higher level implementation is to use JavaBeans-components to add a superstructure for a component type system.

JavaBeans-components model is not "logically closed" in ideology point of view because of the following consideration. The components are supplied in form of classes (implementing abstract data types) to be instantiated, and the classes were designed in accordance with object-oriented *class-based programming paradigm*. When composing their instances in some composing environment (e.g. in the BeanBox), a composite object is created that is not an instance of some composed type; it is just a content of the pre-existing container instance it was built inside (i.e., the content of a BeanBox container instance). In JVM, types of objects may come into existence only by loading byte-codes of the corresponding classes. That composite object can be cloned (serialized / de-serialized, etc.), but its usage in that way corresponds rather to *prototype-based programming paradigm* (we have no class created for it during the composition) [13]. When performing a components composition without its code generation we have to use the composite object as a prototype, thus substituting initial programming paradigm by another. We cannot produce the composite entity of the same nature as we had initially to compose it (without having to use the same technology as we used to produce initial components).

Note though, that in electrical engineering, for example, we can build a functional unit from its components using much more simple technology than technology used to produce them (and it is a matter of integration density). We can draw its scheme as the composed unit type description and put it into production for future reuse. In case we could be able to use microcircuits with more density, we could implement the composed unit in one chip using chip manufacturing technology that we used before for our components manufacturing. But meanwhile we just can use soldering-iron with wires. Roughly to say: that's why Intel develops sophisticated chips while others use to wire them together in more simple manner, placing them wired on PCBs for different purposes.

What we are looking for is a relatively lightweight, dynamic, interactive composition technology to create new composed components without their codes generation.

## 3. TYPES FOR DYNAMIC COMPONENT COMPOSITION

When coding in Java we can only write data type definitions. These type definitions are compiled into class-files with byte-codes that appear to be the runtime types in JVM upon class loading. Inside JVM, the types are represented by objects of type Class that are produced by class-loaders. All that classes are immutable objects (static fields in classes can be mutable, but that style of programming it is not considered as good one). Loaded classes cannot be considered as JavaBeans-component instances: they cannot be created by default-constructor of their class (Class); they are created by some ClassLoader instances having their byte-codes as input.

If we want to cross the boundaries of the runtime type creation ideology of the JVM, we need to define some superstructure over the JVM that has its own notion of object type. Since we are going to have that superstructure as a kind of JavaBeans component model extension and to implement it using our JavaBeans components (in component-based manner), we name it as BeanVM. The type notion in BeanVM should allow different type implementations: both types created from JVM classes loaded by class loaders and types created by our composition procedure designed specifically for that purpose. It means that BeanVM types can be classified as *hardcoded-types* and *composed-types*; the former come into BeanVM from loaded JVM classes, the latter are produced inside BeanVM itself by composition.

Some of the BeanVM types are components (in BeanVM perspective, no matter how they are implemented). For now, we define the types that can be instantiated without any information provided from outside (i.e. from their instantiation context) as components (and we'll try not to mix them with the component instances, as it often happens in JavaBeans related texts). That definition recalls JavaBeans-component definition for JVM that must be instantiable using its default-constructor.

Like JavaBeans-components, BeanVM components can have named property sets with typed property values (i.e., having property value types in terms of BeanVM types). All BeanVM types are instances of the Type type, like all Java classes are instances of the Class class (for short, we omit actual package names here). Type type is implemented in Java by (abstract) class Type providing all type related operations for BeanVM. All BeanVM types are implemented as immutable objects – the information they contain does not change after they have been created. We are intentionally following the class-based object-oriented principles and trying to retain them when performing components composition (in contrast with JavaBeans component model, as it was mentioned above).

To access BeanVM functionality implemented in Java we provide the BeanVM API that we discuss below when needed.

### 3.1. Type representation

Any BeanVM type is represented by an immutable instance of that type implementation java class that is inherited from the abstract class Type and exposes the following type information:

$$type = \{typeName, interfaceType, implementationType\}, \qquad (1)$$

String typeName provides the name of the type; interfaceType and implementationType reflect the type interface and implementation, correspondingly.

We expect to deal with types that were not necessary compiled and, therefore, may have no type-specific methods to be invoked by any method calling mechanism of JVM. The only ports that can be used to communicate with our BeanVM component instances are properties that are supported by special BeanVM API for component instance property access.

An instance of the InterfaceType class contains a set of PropertyType objects describing the interface properties that form the interface of the type:

$$propertyType = \{propertyName, valueType, accessType, defaultValue\}, \qquad (2)$$

String propertyName is the name of a property. The valueType is the fixed BeanVM type that is used for type control when performing new value assignments: each BeanVM type can verify whether the given object belongs to its domain of values. The accessType defines operations to be applicable for the given property. A property can be readable (R), writable (W), bound (B). Indexed property can be indexed-readable (IR) and/or, indexed-writable (IW) as well. The defaultValue for a property is available for properties of the types that are components, only.

The implementationType part of the type provides information on internal type implementation that is different for *hardcoded* and *composed* types.

Any object that our BeanVM can deal with when it is functioning has its BeanVM type, and each BeanVM type can verify whether the given object belongs to the set of values of this type. That is used to implement value type control for property assignments. If an object was instantiated by BeanVM type, it knows that type upon creation. If an object was created by JVM class instantiation and that class defines a component property value type, then that JVM class has been wrapped by the corresponding BeanVM type that delegates the object type check back to its implementations class. BeanVM supports BeanVM type instantiations and property access for their instances; all other activities are performed beyond the scope of its responsibilities (i.e., inside the behaviour logic of the components themselves).

## 3.2. Hardcoded Types

The hardcoded types in BeanVM are types implemented by loaded JVM classes. Hardcoded components are hardcoded types that are components. Here we consider how to represent JVM classes as BeanVM types and how to represent JavaBeans components as BeanVM components.

The source information for hardcoded type definition in BeanVM is provided by corresponding class-object loaded in JVM. We implement a primitive to map Java class to hardcoded type by implementing the following method in Type type implementation class (as part of BeanVM API):

$$public \ static \ Type \ Type.forClass(Class \ someJavaClass); \qquad (3)$$

We have implemented our TypeLoader's hierarchy that mimics that of ClassLoader(s) since each Java class is identified by its' name and the class-loader instance that the loaded class can provide. When implementing our TypeLoader(s), we enhance the possibilities to create types in BeanVM: we can create types not only from loaded classes, as the primitive above does, but in some other ways described later.

We create a type for a given Java class once only, at the first attempt to get it; all subsequent calls will return existing hardcoded type from the type-loaders' (HashMap) table.

The hardcoded type creation procedure is based on reflection mechanism and standard JavaBeans introspection. The set of PropertyType objects is created based on the array of PropertyDescriptor objects that are provided by JavaBeans introspection procedure for the class (JavaBeans introspection can deal with any Java class, not only with JavaBeans-components). The PropertyType object (see (2)) gets the propertyName extracted from PropertyDescriptor object according to the JavaBeans design patterns [4]. The valueType is a BeanVM type created for a class of the property value by means of Type.forClass()-primitive above (3). The property value class and the information to define the property accessType are available in the PropertyDescriptor object as well.

The InterfaceType object with the PropertyType objects array inside can be obtained from any Java class, but we are interesting in JavaBeans-components classes and their property value classes (that we wrap by our types). BeanVM-types for other Java classes are out of interest for the BeanVM.
The ImplementationType object for the hardcoded type is just a wrapper of the source class that implements the type in BeanVM (it reflects the old idea that any problem in Software Engineering can be solved by additional indirection).

The defaultValue in the propertyType (2) is needed in interface type for components only, and it is obtained only from our JavaBeans-components having our specific implementation. For all other (third party) JavaBeans components we provide our hardcoded component-adaptor (that wraps any extraneous JavaBeans-component to be used in BeanVM environment).

Our hardcoded components are JavaBeans components having their implementation class inherited from our Bean class (directly or indirectly). The Bean class provides BeanVM API to the internal implementation of the Bean component instance and its properties. All our hardcoded components are JavaBeans components that have the correspondent BeanVM type instance implementation wrapped inside. The property access methods of our JavaBeans component implementation use BeanVM API and delegate to the wrapped instance implementation.

Here is a code snippet of the Bean class:

```
public class Bean extends BeanVMObject {

final Instance thisInstance; //the instance of the correspondent BeanVM type

public Bean ()    {thisInstance = Type.implementBean (this);}
Bean (Type type) {thisInstance = type.createInstance (this);}
    // …
protected final void initPropertyValue (String propertyName, Object v){…}
public final Object getPropertyValue (String propertyName) {…}
public final void setPropertyValue (String propertyName, Object v){…}
    // …
}
```

The BeanVMObject is an abstract class that forces all BeanVM objects classes to provide their type getter method. Bean class has two constructors: public default constructor and package private constructor with a type argument. The default constructor, that is unavoidably executed when instantiating any Bean class ancestor (any our JavaBeans component), passes this component instance to its implementation factory method - Type.implementBean (this), that returns the instance implementation. The factory method, first, gets the type for this class and, second, lets the type to create the instance implementation. Hence, any Bean class ancestor in its constructor context is ensured that its implementation instance is already created and all its properties are implemented in it as appropriate. The concrete ancestor component is able to initialize its properties with their initial values using initPropertyValue () – method, and use the property value access methods (setPropertyValue (), getPropertyValue (), etc.), that all delegate to the implementation instance.

Note, that initPropertyValue ()-method works only once for a given property during the component type creation, when the component is instantiated for the very first time to collect property default values only (and store them in PropertyType objects (2)). When hardcoded type is created, the initPropertyValue ()-method for its instances has no effect: they are already initialized with their default values when they are created in the implementation instance.

Hardcoded component instance implementation contains only its mutable properties; immutable property values are stored in the PropertyType objects and shared by all instances of the type. The mutability of the property is determined by its accesType: the property is mutable if it is writable or bound (i.e., can change its value externally or internally, notifying about that).
Internal BeanVM API implementations of the property value access methods control the property accessType. All setPropertyValue ()-methods control the property value type. All that control is implemented dynamically, and a kind of RuntimeException is thrown when violation occurs. To speed up the property access we provide methods in BeanVM API that translate the propertyName into an index of its PropertyType object in the component interface type along with the variants of property access methods that use the index instead of the propertyName. The concrete hardcoded component implementation can get its property indices in static initializer of its implementation class and use them when implementing its property access methods according with JavaBeans design patterns. In JVM perspective, all property values are stored internally in the BeanVM memory cells allocated for them and declared using the root of JVM types hierarchy (as java.lang.Object). That requires an explicit cast to the property value type to be added when coding the concrete property getter using our getPropertyValue()-methods that return Object (some overhead with Java primitive types could be minimized by providing specific BeanVM API methods for them, but we omit these details here).

Note that our hardcoded components are JavaBeans components that can be manipulated visually (by definition) in a builder tool like the BeanBox. It means that we keep all advantages of the JavaBeans component model, and we can use our hardcoded components, e.g., like we did before when implementing our 3D modelling framework by means of JavaBeans [6]. We could build a composite model from JavaBeans component instances and observe the model behaviour, having placed it inside some predefined component container instance. But without generating byte-codes and loading them into the JVM, we could not create new composed component and place it into the ToolBox of the BeanBox (to say that in visual terms), while all the components we used to create it are already there. Now our goal is to make it possible: we have to deal with composed types and be able to extend the builder tool (like the BeanBox), accordingly, to deal with them as well.

### 3.3. Composed Types

Composed type in BeanVM is a type that is not implemented by some existing java class loaded in JVM, but is defined by a special data structure describing its composition from its composing parts. It has interface definition and internal (hidden) implementation that is defined using other components as the building blocks.

When defining composed types we comply with the object-oriented class-based paradigm: our type definition is immutable structure that is able to create instances of that type instead of cloning them. The composed type that is instantiable without any arguments provided for it is composed component.

Like any type in BeanVM, composed type has its type name, the interfaceType and the implementationType (1). The interface type for the composed type is represented the same way as for a hardcoded type and is described by the array of PropertyType objects (2). But the implementationType is entirely different: it is composed by its *composing types* that are organized in a directed acyclic graph representing implementationType definition. Composing type is a context-dependent refinement of a composing component that describes how its usage in the given context of the composed type implementation (i.e., as the composing graph node) differs from its own (context-less) component definition. These differences are defined in terms of some property types' attributes that can be context-dependent: access rights limitation, property value, and, possibly, property implementation sharing (that we discuss below shortly).

Public BeanVM API for property access works equally for any instance of any BeanVM type - no matter whether it is a hardcoded type or a composed type. External communication with a composed component instance is performed using its interface properties access by means of BeanVM API mentioned above for hardcoded type instances: any instance of a composed component is implemented as our Bean class ancestor instance that gets its composed type as an argument of the package private constructor (see Bean class code snippet above) called when the composed type instantiation is performed.

When creating an instance of a composed type, we create the instance internal implementation that, in this case, includes not only the mutable properties cells (as it is done for hardcoded component instances), but the implementationType instance as well. That implementationType instance is created as the result of composing graph traversal procedure where composing types that are met during the traversal are instantiated to provide the composed type instance implementation.

For each instance of the composed type, the instance internal implementation should be able to communicate with the instance external environment through the interface properties. The composed type instance behaviour is implemented by its composing components instances that express their behaviour by their property value changes – as any our component instance does. Hence, to link the interface with internal implementation of the composed type instance, we need to link some interface properties to some properties of internal composing components instances. These composed type interface and implementation properties links can be organized basically in two different ways: by bidirectional property bindings and by interface property implementation sharing. The second way is less expensive from the efficiency point of view, since there is no need to transfer property values.

Since we are supporting class-based object-oriented approach, we delegate our BeanVM API operations to the corresponding type object that implements them using the concrete instance reference (an internal instance implementation). When implementing an interface property access, we delegate to the interface property type that knows how to find and access the property value having the internal instance implementation. To implement that, we have PropertyType class sub-classed according property implementation categories: Immutable, Mutable, Bound and External. Each category (that is the PropertyType subclass) is responsible for corresponding property implementation and access. All mutable (and bound) property values are stored in an array of objects that is allocated by internal instance implementation factory. All immutable property values are stored in their types and are shared by design. PropertyType.External is used to share the enclosing type interface property with the given one. The PropertyType.External delegates the property access to the given property of the enclosing instance of the composed type. That enclosing instance is known as the composing type instantiation context during the traversal procedure of the composed type implementation graph.

We have mentioned previously that our components are instantiated without passing any arguments for their default constructor, as it was stated in the JavaBeans component definition. All our hardcoded components are JavaBeans components that are instantiated by their default constructors (without passing any context-related information). We do not use the BeanContext-related extensions of the JavaBeans Specification to inject context-related information later, as well. While JVM provides some tricky way to get some information about constructor calling context using invocation stack trace, we do not rely on its usage. Instead, we use the fact that we are working in BeanVM (instead of JVM), that implements its own component instantiation primitive, createInstance (). That primitive implementation calls Java default constructors of all our components internally, and that default constructor (in base Bean class) delegates the internal instance implementation back, to the BeanVM implemented instance factory method (see Bean class code snippet above). That factory method is able to use BeanVM primitive invocation stack to get the instantiation context (if it exists) and pass it to the internal instance implementation.

All the information needed to implement and access composed component instances is contained in their composed types implemented as immutable data structures that expose property value getters for reflection purposes (as implemented by abstract Type class and its type-specific subclasses). Next we'll describe how that structure can be created dynamically.

### 3.4.    Prototypes

The natural way to create an instance of immutable data structure representing a composed type is to use its builder object that is mutable and can collect all relevant information to be provided for immutable object initialization (i.e., to use the Builder design pattern [19]).

We create the builder object in component-based manner and use our specific hardcoded components to compose (using their instances) a *prototype* object that can serve as a source for creating an immutable composed type. These specific hardcoded components can be considered as meta-components (since they are components to construct components).

The prototype object is an instance of the hardcoded component Prototype having properties "name", "interfacePrototype" and "implementationPrototype" (values of these properties will be used to define properties for the composed type created from that prototype). The "interfacePrototype" property can refer to an instance of hardcoded component, InterfacePrototype, which is used to build the interfaceType. Property "implementationPrototype" refers to an instance of hardcoded component, ImplementationPrototype, used to compose the prototype implementation object.

The prototype to type transformation is performed by BeanVM API primitive (here, for short, we omit the details concerning nested types that are enclosed in outer type implementation):

public static Type createFromPrototype (Prototype prototype);                    (4)

The new composed type is returned in case when nothing prevents it from being happened (i.e., there are no namespace conflicts and other validation faults).

For instance, we can create BeanVM type just having set the name only (i.e. having empty interface and empty implementation), that is similar to an empty JVM class (e.g., class Classname{}). Type with some non-empty interface part present, but with an empty implementation part can be used to instantiate its property set without any internal behaviour linked with them (like a structure or record). Type with no interface part can represent a separate executable entity type (e.g., "a scene" with its separate behaviour).

The instance of the InterfacePrototype component exposes the set of property prototypes. The property prototypes serve two goals: 1) they are used as exposed handles to control the prototype instance behaviour, and 2) they provide source information to create the property types during the prototype to type transformation.

The instance of the ImplementationPrototype component exposes the set of component instances that compose the prototype object implementation – the set of composing prototypes. The compound prototype object can be built by defining graphs of two kinds: reference graph and events graph. The reference graph is defined by using some object as a property value (or as an element of indexed property value) of another. When an object is used as a property value of (i.e., is referenced by) several other objects, then it is shared by them. The events graph is defined by component instances that bind the source bound property to the target property to propagate the property change events (in correspondence with JavaBeans design patterns).

One of the main principal issues when defining composed components is defining a way to separate the interface of the component from its implementation while providing the way for them to intercommunicate. We define the interface in terms of properties that are prototyped using typed variables, i.e. objects having "value'-property with a given value type. We can use any BeanVM type as a value type for our typed variables (and as property value type after the prototype to type transformation). In fact, that is one of the two existing use cases for BeanVM types; another is BeanVM type instantiation. Having a type to be used as a value type, BeanVM creates (synthesizes) the BeanVM type for the typed variable automatically, assigning the synthesized name for that synthetic type and granting the full access rights to operate with its instances (i.e., rights to read, write, and bind their property "value"). In case the value type is an array type, we create BeanVM type for indexed typed variables that will serve as indexed property prototypes (having indexed access provided). That approach is similar with JVM array classes' creation and support (based on a given class of the array elements).

Property prototypes, implemented using typed variables and exposed through the interface prototype, should be linked with property prototypes of some composing prototypes inside the prototype implementation. It can be done either by means of event routing graph (using bidirectional event routing for each link), or by sharing the property prototypes (by means of reference graph). We use the latter approach (the more effective one) and provide support for the property prototypes sharing, that corresponds to the similar approach for composed type instances implementation.

When a component is instantiated inside a prototype container, it gets its *prototype-oriented internal implementation* from the instance implementation factory (in contrast with the component instantiation in other container contexts). We use that instance as a composing prototype. The prototype-oriented implementation works in prototype-based style: it does not delegate the property access methods to the type of the instance (since it does not exist yet), but implements them using the prototype-oriented instance implementation itself. In that internal implementation, a composing prototype instance is represented with an array of property prototypes (in contrast with an array of property values, as it is implemented for component instances created outside the prototype container context). That additional indirection supports sharing the property prototypes of the interface prototype by composing prototypes instances inside the prototype implementation part (in case the value types are compatible). In principle, that sharing is similar with reusing the component instances inside the reference graph of the prototype implementation.

Each property prototype is supplied with access prototype instance that can be used to narrow the property prototype access rights by denying some of them for the given usage context. After the prototype to type transformation, narrowed access rights will be stored as accessType in the correspondent property type, as was mentioned in (2), Section 3.1.

The composed prototype object is tuneable and operational. Its' composition can be performed by visual manipulations with the correspondent tools support. When it is done, it can be used to produce the immutable BeanVM type. During the prototype to type transformation all prototypes are used as sources to create the corresponding types: property prototypes are transformed into property types with their access types created from the access prototypes, interface prototype is transformed into the interface type, composing prototypes are transformed into composing types, altogether transformed into the implementation type, and finally resulting in the composed type, or component, produced.

That newly created composed component can be instantiated as any other component –either using more efficient class-based object-oriented internal implementation, or using the more flexible prototype-oriented internal implementation (having been instantiated as a composing prototype in the prototype container).

Composed component can be serialized and de-serialized (e.g., using some text format). To read the composed type from a text file by BeanVM TypeLoader, we provide Type.forName (String typeName) primitive that loads the type by its name like JVM Class.forName (String classsName) loads classes. When looking for the source of the type by its name, we first try to load hardcoded type with the given name (using Class.forName ()), if it exists, then the serialized type file to be parsed. The parser, essentially, reads into a prototype object and performs the prototype to type transformation.

## 4.  SAMPLE APPLICATIONS

The VRML [7] and its successor X3D [20] Standards define sets of elements that constitute a scene to be depicted using 3D and/or 2D graphics. The scene model in memory is represented by the directed acyclic graph (DAG), consisting of node instances whose types are predefined according to that standards. The DAG is formed by node instances containing typed 'fields' whose values are other node instances. In that way, a node instance can be referenced by others (i.e., reused or shared by them) provided there are not cycles in the reference graph.

Values of nodes fields, in common, define the (scene) model state, and can vary, in event-based manner, while event propagations are performed among them. For all the base predefined nodes, their field value types are statically known and defined by the Standards. Directed acyclic graph can be traversed with the result of earning some context-specific information, which, in combination with nodes field values, defines the visual presentation of the model that is rendered. The model behaviour is defined by changes that happen in the node graph during events propagation, with events carrying data on their field value changes from node source to the node target, and by reactions in receiving nodes, that, in their turn, can change their state and fire events. VRML and X3D Standards define base node types with their field types (and their meanings/semantics), the constructs to define event routing graph, standard scene access interfaces and so called 'profiles' – sets of independently distributable standardized functional components to support various (extendable) modelling abilities.

While the Standards do not expect their Java-implementation (and up to date they were not implemented in Java properly, despite of some attempts), their implementation using JavaBeans component model appears to be pretty natural; moreover, using JavaBeans-component model we can avoid some limitations of the Standards, that were not initially designed to use neither dynamic abilities of the Java platform nor component model for it. JavaBeans components usage to implement subsets of the Standards with some additional features were discussed in [6], that shows how parallel compositions, behaviours and 3D-visualization of the models can be organized in standard JavaBeans-container (using BeanBox) along with standard JavaBeans-components.

Component-based approach provides for enhanced flexibility when defining the content of standards specifications. Profiles can be considered as component sets (defined by their authors), and supplied without having been defined by some standardization authority. Similarly, the component sets can be extendable in natural way (that is the matter of deployment packaging); field (or property) value types can be defined by component authors, without limitations, etc. What should be standardised actually – is the component-based approach to the development and the corresponding component model itself.

Using the presented approach and component model, we can implement user-defined nodes definitions that are described by the PROTO construction found in VRML and X3D, in class-based object-oriented manner. It was not possible using JavaBeans component model, as we have discussed above.

The PROTO construction in VRML (or X3D) essentially provides the language features to describe new, user-defined node type, having supplied its name, its interface in terms of its fields (as that is done for the base, predefined node types) and internal implementation, that is similar to a scene graph. Both scene graph and PROTO implementation graph can contain nodes of any type – both predefined and user defined – provided they are defined prior to their usage in a graph by means of PROTO construction. Node sharing is defined using DEF/USE construct where DEF provides a name for a node instance, and USE refers to it in a field value. Fields from the interface of a PROTO are bound to fields of its internal implementation nodes using special VRML syntax construction – "IS". (In our model that separate construction is not needed – it is essentially the same as DEF/USE construction to share interface property).

In practice the PROTO construction is implemented as a prototype, as its name implies, to be cloned, or just as macros definition to be substituted by parser. The whole scene description is not considered as a type definition to be instantiated; instead, it is just a prototype instance serialized in VRML or XML (in X3D) formats. All existing VRML/X3D implementations are developed corresponding to the prototype-based ideology.

Using the proposed component model we expect to simplify and optimize software systems for the subject areas, where declarative languages (like VRML or X3D) are used to describe 3D-models with event-driven behaviours, and we will be able to support new abilities of these systems to benefit from.

The ability to build a flexible prototype object from components that can be transformed into composed instantiable component with context-dependent optimization may be useful in various applications. For example, that ability directly corresponds to the wireless sensor network commissioning task [21]. These networks can be described using a graph of nodes communicating by radio in event-driven manner, and are built from standard components to be tuned for a given application and environment. The concrete ad hoc network prototype (or a model) can be designed, then that typical network component can be created, and its instances, containing concrete network nodes settings, can be used for commissioning (e.g., over the air). That kind of modelling does not require 3D/2D visualization, while can benefit from it (similarly, VRML/X3D engines may be used not only for visualization tasks).

## 5. RELATED WORKS

Component-oriented programming and its usage in software development formed a component-based software engineering – a branch of software engineering that has its long term history, valuable results and issues to be solved. There are many publications attempting to provide definitions for a component (we have referred to three most frequently cited [9, 10, 11]), investigating different component models with their taxonomies [2], and considering various aspects of component-based software development for different application areas [10, 11, 15]. Detailed overview of that works is beyond the scope of this paper.

We are concerned with issues we know from experience earned while developing component-based software system to implement 3D modelling and visualization [6]. This is a large area as well, and we will narrow our scope by platform in use, since there are popular platform dependent component models we have to leave aside, e.g. [22].

The most widely known 3D modelling software for Java platform is probably Java 3D [23] library, initially developed by Sun, then by the open source community. That heavy-weight library was not designed with component-oriented approach in mind, and attempts to implement VRML browser using Java 3D library were not finished. The library was used in X3D Standard development by web3d.org community [24], but currently Java 3D library is deprecated.

There are several 3D modelling and visualization tools for Java platform developed by commercial companies and universities, e.g. [25, 26], but they do not explicitly use a component-based approach or use a proprietary component model by their own [27].

The PtolemyII project [28], having its long pre-Java history and covering, among others, the application area of our interest, has been moved to Java platform and use its component model reflected in Moml [29] specification. As far as it can be seen in publications, instantiation of declaratively defined composed type in it is implemented by cloning (i.e. in prototype-based manner).

At the time of highest VRML popularity, there were publications on extending it in object-oriented manner [30, 31]. Now we observe new wave of interest to these declarative languages that can be seen in new VRML/X3D compliant product – Instant Reality [32]. But that product does not extend the standards in component-oriented direction.

Both VRML [7] and X3D [20] standards specify Java authoring interfaces (for external model access and for internal scripting in Java). All that specifications have no relation with component models for Java-platform, while that functionality (actually, much more) could be readily provided by component-oriented design.

There are some works on component-based software evolution, e.g. [33, 34], but they are based on code generation. We can consider code generation as a feasible way to translate BeanVM components into JavaBeans components (that is similar to the Java just-in-time compiler translating from virtual machine level to executing machine level).

## 6. CONCLUSION

Both hardware and software architecture histories demonstrate the evolution with increasing dynamics abilities. Java platform, as the most popular software development environment, and JavaBeans component model, the most popular component model for that platform, provide the means to evolve in that direction as well. In this paper we have proposed an approach and a component model that demonstrate prototype-based and class-based programming paradigms interrelations. The prototype is a flexible, mutable object sample that is built, lives and evolves in component-based manner; when it gets some desirable state of evolution, or just eventually by some stimulus, its genetic code is extracted into the type and saved for reuse by next generations - that reminds an everyman untaught view on genetics in the natural reality.

The component model proposed is a kind of dynamic extension for the JavaBeans component model, with the extension supported by specific JavaBeans components. Correspondingly, as for JavaBeans component model, we need the extended builder tool that utilize and demonstrate extended abilities of the component model. Developing that tool is the goal of our future work.

**Author**

**Efim M.Grinkrug** (egrinkrug@hse.ru)

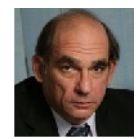

currently serves as the Associate Professor at the Software Engineering Department of the National Research University Higher Scholl of Economics, Moscow, Russia. Formerly worked as the CTO of MeshNetics (ZigBee company), as the expert at ParallelGraphics (VRML company) and as Operating Systems developer at the R&D Institute for the Calculating Complexes named after M.Kartzev, Moscow, Russia.